\documentstyle[12pt,aasms4]{article}
\received{(revised) \today}
\accepted{ }

\newcommand{\co}{$^{12}$CO\ }
\newcommand{\coc}{$^{12}$CO}
\newcommand{\tco}{$^{13}$CO\ }
\newcommand{\tcoc}{$^{13}$CO}

\newcommand{\kms}{ \hbox{km~s$^{-1}$}\ }
\newcommand{\kmsc}{ \hbox{km~s$^{-1}$}}

\begin{document}
\title{The Bell Laboratories $^{13}$CO Survey:
\\Longitude-Velocity Maps}
\author{Youngung Lee\altaffilmark{1,2}, Antony
A. Stark\altaffilmark{1}, Hyun-Goo Kim\altaffilmark{2}, and Daesik
Moon\altaffilmark{3},} 

\altaffiltext{1}{Harvard-Smithsonian Center for Astrophysics, 60
Garden Street, MS-78, Cambridge, MA 02138; ylee@cfa.harvard.edu}

\altaffiltext{2}{Korea Astronomy Observatory, Taeduk Radio Astronomy
Observatory, Whaam-dong San 61-1, Yusong-gu, Taejon 305-348, Korea;
yulee@trao.re.kr}

\altaffiltext{3}{Cornell University, Ithaca, NY 14853}

\begin{abstract}
A survey is presented of the galactic plane in the $J=1-0$ transition
of $^{13}$CO. About 73,000 spectra were obtained with the 7 m
telescope at Bell Laboratories over a ten-year period.  
The coverage of the survey is ($l, b$) =
($-5^{\circ}$ to $117^{\circ}, -1^{\circ}$ to $+1^{\circ}$),
or 244 square degrees,
with a grid spacing of $3'$ for $|b| <
0^{\circ}\!.5$, and a grid spacing of $6'$ for $|b| > 0^{\circ}\!.5$.
The data presented here have been resampled onto a $3'$ grid.
For 0.68 km s$^{-1}$ channels, the rms noise level of the survey is
0.1 K on the $T_R^*$ scale.  The raw data have been transformed into FITS
format, and all the reduction processes, such as correcting for
emission in the reference positions, baseline removal and
interpolation were conducted within IRAF using the FCRAO task package
and additional programs. The reduced data are presented here in the
form of longitude-velocity color maps at each latitude.  These data
allow identification and classification of molecular clouds with masses
in excess of $\sim 10^3$ solar masses throughout the first quadrant of
the Galaxy.  Spiral structure is manifested by the locations of the
largest and brightest molecular clouds.  

\end{abstract}


\section{INTRODUCTION}

Surveys of carbon monoxide (CO) emission lines are the primary
means for gathering
information about the distribution and kinematics of molecular
material in the Galaxy.  The observer, when embarking on a CO survey,
must choose the survey parameters with an eye to the astrophysical
problem to be attacked, because millimeter-wave telescopes have
limitations in observing speed, sensitivity, and resolution.  Rapid
improvements in technology have allowed ever larger and more
finely-detailed surveys, but the current state of the art is not yet
able to produce detailed, low-noise maps of every molecular cloud in
the Milky Way.  Combes (1991) has reviewed sixteen CO surveys and
discussed the results.  Early surveys, such as Schwartz et al. (1973),
Burton and Gordon (1978), and Solomon et al. (1979) determined the
large-scale distribution of CO in the Milky Way.  The Columbia Mini
and Southern Mini surveys (Dame et al. 1987, Bronfman et al. 1988), in
particular, measured the CO emissivity of the whole Galaxy and
determined its distribution on scales larger than $\sim 50$ pc.  Later
surveys have concentrated on molecular cloud properties and the
relationship between molecular clouds and galactic structure.  The
Bell Laboratories \tco Survey presented here was designed to identify
most of the molecular clouds in the first quadrant of the Galaxy and
provide some information about their size, shape, and velocity.  It differs
from previous surveys in the accuracy with which clouds can be identified
and the molecular line emissivity parceled out among them.
These data will
be used to approach the problem of spiral structure and its
effect on molecular clouds and star formation.

Several statistical
studies (Scoville et al. 1987; Solomon et al. 1987; Sodroski 1991) of
the giant molecular clouds (GMCs) identified in survey data
have led to significant results on the kinematics of disk structure,
locations and extent of spiral arms, and the mass and size spectra of
the GMCs.  These analyses were based on a 
simple methodology for recognizing molecular clouds in
survey data.  In these studies, a molecular cloud is defined as a
region in $(l, b, v)$ survey space where the \co line emission
exceeds a particular threshold brightness temperature.
The tendency for \co lines from different clouds to blend together 
because of spatial and velocity
crowding and optical depth effects has led to the
choice of a high value for that threshold.
Scientifically important inaccuracies in our quantitative
understanding of the distribution of molecular gas may result,  
especially in crowded regions such as the Molecular Ring.
Small clouds may be missed entirely because even their central
brightnesses may not exceed the threshold, while large clouds
are sometimes spurious blends of objects at different distances. 
Significant crowding of molecular material occurs throughout
the first and fourth quadrants of the galaxy, especially on the
locus of tangent points. 
Along any line of sight toward the inner Galaxy, 
the point on the line closest to the Galactic Center is the
tangent point, and clouds near this point have the largest velocity and the
most velocity crowding.  Clouds at the tangent point are, however,
particularly interesting
because they are at an inflection point in the velocity-distance
relation and therefore have unambiguously determined distances.

The \co line becomes optically thick towards most molecular clouds.
The isotopically-substituted \tco molecule is a more accurate tracer
of molecular column density, because its optical depth is a factor of
$\sim 50$ times smaller than that of \coc.  Data obtained in the
\tco(1-0) emission line with a large spatial coverage and a high
resolution can be used to quantitatively study the molecular material
near young stellar objects (YSOs), since the more optically thin
emission traces dense regions where star formation takes place.  While
the \tco lines are several times weaker than \coc, its smaller
opacity, relatively high abundance ([\tcoc]/[H$_2$] $\sim 10^{-6}$ in
the solar neighborhood---Frerking et al. 1982), and the relatively
greater transparency of the atmosphere at 110 GHz make \tco the
species of choice for surveying molecular gas in crowded regions.
Surveys conducted in \tco can be used to separate and catalog clouds
better than \co surveys.

The primary disadvantage of \tco surveys is that they are less sensitive
to weak CO emission than are \co surveys.  Weak \co emission has been
detected in small interstellar clouds (e.g. Magnani, Blitz, and Mundy 1985) 
and in a thick molecular layer associated with the whole galactic
disk (Dame \& Thaddeus 1994).  Such material has a \tco / \co 
line intensity ratio which is smaller than that of most molecular material,
and the \tco emission is therefore more likely to be buried in
noise.  The low noise level of the present survey partially compensates 
for this effect.   In the region covered by the present survey,
almost all clouds seen in \co surveys are also seen here.
 
Several \tco surveys were completed in the 1980s (Liszt et al.  1984;
Jacq et al. 1988; Bronfman et al.  1989).  These are, however, all
one-dimensional---limited to the galactic mid-plane ($b = 0^{\circ}$).
Aside from the Bell Laboratories \tco survey, systematic
two-dimensional \tco surveys of the galactic plane have been conducted
only toward limited regions: Nagoya University's \tco map made with
their 4m telescope ($2'\!.7$ beamwidth) with a spatial sampling of $8'$
toward the Cygnus region (Dobashi et al.  1994), Taurus (Mizuno et
al. 1995), the Cepheus and Cassiopeia region (Yonekura et al. 1997),
and most recently, the Antigalactic Center region (Kawamura et
al. 1998).  There are plans to survey the whole galactic plane with
the two Nagoya 4 m telescopes (Nagoya University in Japan and Las
Campanas Observatory in Chile), and a significant portion has been
completed (Tachihara et al. 1997).

In this paper, we present the results of a \tco survey covering 
a third of the galactic plane at a spatial sampling of $3'$
using the 7 m telescope at Bell Laboratories.  The
Bell Laboratories \tco survey was designed to detect and distinguish molecular
clouds in the first quadrant larger than about $10^3$ solar masses.
The survey has enough resolution to locate star-forming regions
within large clouds.
It does not, however, provide much detail of the internal
structure of the clouds.  The large longitude coverage allows the study
of giant molecular
clouds in a galactic context, while the sampling grid size allows some
indication of their internal structure.
In \S 2 is a description of the survey parameters,
telecope and receiver systems, observing techniques, and reduction
procedure. In \S 3 the data are shown as longitude-velocity
color maps, one at each latitude, and the principal characteristics of the
survey data are discussed.

\section{The Survey}

\subsection{Telescope and Receiver System}

The data were obtained with the 7 m off-axis
Cassegrain radio telescope located at Bell Laboratories, Crawford Hill
in Holmdel, New Jersey (Chu et al. 1978)
between 1978 and 1992. The beamwidth (FWHM) of the
telescope is $103''$ at the frequency of \tco $J=1-0$ line (110.20913
GHz).  A Schottky-diode receiver system developed by R. A. Linke 
was used for the early data;
after 1982, Superconductor-Insulator-Superconductor (SIS) mixer
receivers developed by R.A. Linke and A. A. Stark were used. The
superconducting junctions were fabricated by R. E.  Miller and
A. A. Stark at Bell Laboratories, Murray Hill.  Single-sideband (SSB)
operation was achieved using a Fabry-Perot filter in reflection which
had 11 to 15 dB rejection of the image sideband.  The unwanted image
sideband transmitted through the Fabry-Perot filter was terminated in
a millimeter-wave load with a 28 K radiation temperature.  The
receiver was located at a non-tilting Nasmyth focus in a housing
located on the azimuth structure.  The receiver was calibrated by
chopping between an ambient temperature load and a liquid
nitrogen-cooled load located near the Cassegrain focus.  Single
sideband receiver temperature of the receiver systems ranged from 76
to 360 K during the years of observation, with typical values around
120 K.  Atmospheric sky brightness was measured several times each
hour by chopping between the 77 K Cassegrain cold load and the sky.
The correction for atmospheric opacity was then calculated using a
plane-parallel atmospheric model.  Typical SSB 
atmosphere-corrected system
temperatures were $T_{sys}^{*} \sim $ 600 K.  Observations were not
made when $T_{sys}^{*}$ exceeded 1000 K.  The 7 m antenna has a
main beam efficiency of 89\% at 100 GHz; another 5\% of the
detected power comes from the antenna sidelobes and about 6\% comes
from the ground pickup. The spectra were obtained with a 256 channel
filterbank consisting of filters with a 250 kHz spacing and a 512
channel filterbank consisting of filters with a 1 MHz spacing.  The
individual filters in these filterbanks have rectangular bandpasses
with little overlap between adjacent filters.  The frequency domain is
therefore slightly undersampled.

\subsection{Spatial Coverage and Sampling}

The range of the survey is ($l, b$) = ($-5^{\circ}$ to $117^{\circ}, 
-1^{\circ}$ to $+1^{\circ}$), comprising about 244 square degrees.
Table 1 lists the regions observed and the data sampling within those
regions.  In Figure 1, the grid spacing and beamsize of the Bell
Laboratories \tco survey is compared with those of other surveys: the
Columbia \co galactic plane survey and University of
Massachusetts-Stony Brook \co first
quadrant galactic plane survey (UMass-SB) 
are shown in thin solid-line circles,
and Nagoya University \tco survey and Bell Laboratories \tco surveys
are represented in thick solid-line circles on the $3'$-grid.  
A total of 73,000 positions were observed.

\subsection{Observations}

All observations are position-switched data.  Each spectrum is a
weighted average of observations at its ``on" position and two
``off\," positions.  The Galactic Center data are 
described by Bally et al. (1987).  The data with $|l| > 5^{\circ}$
were obtained with a multiple-map position switching technique.  This
means that during the data acquisition, observations of five ``on" positions were
temporally interspersed with observations of two ``off\," positions, and 
the spectra from all
five ``on" positions use the exact same ``off\," measurements.
Each of the two ``off\," positions were observed for
${\sqrt 5}/2$ times longer than the individual ``on" positions in order
to
optimize the noise level and observing speed.  
The multiple-map technique increases
the speed of the observations, but has the undesirable effect that the
noise in the five resulting spectra is somewhat correlated: the ``on"
data for each position (including associated noise) are unique,
but the ``off\," data (including associated noise) are the same
for all five spectra.  The five ``on" positions are separated by
$0.2^{\circ}$, so positions with correlated noise have three other
grid points with totally different noise between them.  

The galactic
coordinates of each set of five points are given by
$$l_{\mathrm on} = L + M \cdot 0.2^{\circ} + \Delta l ,$$ $$b = \Delta b ,$$
where $0^{\circ} \le \Delta l <  0.2^{\circ} , $ 
$-1^{\circ} \le \Delta b \le + 1^{\circ}$ in $0.05^{\circ}$
increments, $M$ varies from 0 to 4 during the 
course of a single multiple map position switch, 
and $L$ varies from $10^{\circ}$ to
$117^{\circ}$ in $1^{\circ}$ increments during the course of a day's observations.  
$\Delta l$ and $\Delta b$ were fixed for each day's observations.  The
galactic coordinates of the reference positions are given by
$$l_{\mathrm off} = L + 0.5^{\circ} ,$$ 
$$b_{\mathrm off} = \pm 3^{\circ} .$$
There are, therefore, only 216 ``off\," positions in the survey.  These
``off\," positions are not necessarily free of emission.  The method 
by which the spectra were corrected
for emission in the off or reference position is described in
\S 2.4.

The typical on-source integration time was  $8.3 \, \mathrm{s}$ for each of
the five ``on" positions and $9.3 \, \mathrm{s}$ 
for each of the two ``off\," positions,
for a total position switch cycle of $60 \, \mathrm{s}$.  
This minute-long cycle was 
repeated 6 to 40 times (depending on the opacity of the sky)
so that the root-mean-square noise in each spectrum, after
calibration, is 0.1 K in 250 kHz channels.  In good weather, it was possible to complete 
a full set of observed points having particular values of $\Delta l$ and
$\Delta b$ (that is, all the points with $M$ and $L$ ranging over their full 
set of allowed values) in one day.  In poor weather the galactic plane would set
before a full set of points could be obtained, and the undone points
would be rescheduled for a later day.

Receiver calibration was accomplished by chopping between an ambient
temperature load and a load filled with boiling liquid nitrogen.  Sky
brightness measurements were accomplished by chopping between the
liquid-nitrogen load and the sky.  These measurements were made at
roughly 20 minute intervals.  
All antenna temperatures quoted here are
corrected for atmospheric extinction and for the forward spillover and
scattering losses of the antenna ($\eta_{\mathrm{fss}}$ = 0.91), 
and are therefore on the $T_R^*$ brightness 
temperature scale as defined by Kutner and
Ulich (1981).  
In order to convert to a $T_{\mathrm{mb}}$ brightness temperature
scale, the data as presented here should be multiplied by $1.02 \pm 0.02$.

These data were obtained during hundreds of days of observing over the
ten-year period.  Observations were made by A. A. Stark, R. W. Wilson,
J. Bally, M. Pound, G. R. Knapp, and W. D. Langer.
Some results of this survey have been presented by Stark (1979, 1983),
Stark et al. (1983, 1987, 1988) and Bally et al. (1987, 1988).
Advances in computation and the recent availability of powerful data
reduction and display packages now permit the full data set to be
re-reduced and presented here.

\subsection{Data Reduction}

The raw OBS-format spectra were calibrated and a linear fit to the
baseline was removed in the Bell Labs data reduction program COMB.
The resulting COMB stacks were written out as 3-dimensional ($v, l,
b$) FITS format data cubes which can be manipulated with the standard
data reduction packages IRAF and AIPS.  For the region $ l >
5^{\circ}$, the COMB stacks data were written onto a
(partially-filled) $3'$-grid in ($l,b$), and a 0.68 \kms (one
filter-bank channel width) grid in velocity.
The size of the data cube outside the Galactic Center region is
($v,l,b$)=(364,2241,41) pixels.  Unfilled positions on the $3'$-grid
were filled in by linear interpolation over the nearest four positions using
IRAF and SAOimage.  We used the FCRAO task package and developed
scripts within IRAF to interpolate and reduce the data.
A second-order baseline fit was subtracted from the spectra.
Bad channels caused by the malfunction of individual filters in the
filterbank were identified in the FITS image and fixed by
interpolating adjacent good data over the bad pixels.

As described above, the ``off\," positions used for the position
switch are not necessarily clean of emission.  About 2\% of the raw
spectra were contaminated with negative features resulting from
emission in the reference positions.  Essentially all of this emission
is within the velocity range $-5$ to 5 \kms and almost certainly originates
in dark clouds near the Sun. The strongest of these features was less
than 0.5 K.  Since each ``off\," position was used for several hundred
``on" positions at the same value of $L$, it is likely that at least
one ``on" position is free of emission at the velocity at which the
``off\," is contaminated.  The ``on" position spectra sharing the same
$L$ were therefore searched for the biggest negative feature.  Data in
a small range of velocities around this biggest negative feature were
extracted from that spectrum.  These extracted data were then taken to
represent the emission in the ``off\," position, and it
was subtracted from all of the ``on" positions sharing the same value
of $L$.  About 0.03\% of the pixels were modified by this method.  A
comparison with the UMass-SB \co database shows that these
recovered pixels do not show a systematically different ratio of \co
to \tco.  

After removing the effects of the contaminating emissions in the off
positions, the database was combined with the
Galactic Center survey FITS data cube (Bally et al. 1987).  The
Galactic Center data were interpolated into a $3'$ grid and a $1\, \kmsc$
grid.  
Thus the final database including the Galactic Center
region is ($v,l,b$) = (501,2441,41) in pixel units and the velocity
range is $-250$ to $+250$ \kmsc, longitude is $-5^{\circ}$ to
$117^{\circ}$, and latitude is $-1^{\circ}$ to $1^{\circ}$.

A histogram of the number of {\it l, b,} and {\it v} pixels versus
brightness 
temperature of the \tco data  ($l > 5^{\circ}$ with 0.68 \kms
resolution) except the 
Galactic Center region is shown in Figure 2 in linear and logarithmic 
scales, and was obtained by binning the data into 0.02 K-wide bins.
The brightness temperatures ($T_R^*$) in the plot range from $-2.5$ to 6 K.  A
few pixels in the survey are as hot as 17.4 K. Only pixels from the
observed ranges of position ($ l > 5^{\circ}$) and velocity listed in
Table 1 are included in this histogram.  

To gain insight about the true distribution of brightness temperatures
if there were no 
telescope or receiver noise, a parametric model was made of the
distribution function.  Suppose the true, noise-free distribution
of brightness temperatures is a power-law in 
brightness temperature, $T$, plus a
delta function at zero temperature:
$$N(T) \, {\mathrm{d}}T = N_{\mathrm{total}} [ f_z \, \delta (0 \, {\mathrm{K}}) + 
(1 - f_z)(\alpha - 1)C^{\alpha-1}
(T + C)^{-\alpha} ] \, {\mathrm{d}}T ~~~~,$$
where $N_{\mathrm{total}} = 23,\! 392,\! 017$ is the total number of pixels
in the survey, $f_z$ is the fraction of pixels that have no
emission, $C$ is a parameter of the fit, and $\alpha$ is the
power law.
The observed distribution of brightness temperatures is given by the convolution
of this function with a Gaussian representing the noise:
$$N_{\mathrm{obs}} (T) \, {\mathrm{d}} T = 
\int N (T - T') \, {\mathrm{d}}T \, {{1} \over{\sigma \sqrt{2 \pi} }} \, 
{\mathrm{exp}}(- T' \, ^2/{2 \sigma^2} ) \, {\mathrm{d}} T' ~~~,$$
and this function can be used to model the actual distribution of
observed brightness temperatures.
A $\chi^2$ fit for the parameters of the model yields $\sigma = 0.1065\pm0.0031$,
$f_z = 0.884 \pm 0.015$, $C = 0.768 \, \mathrm{K} \pm 0.105 \, \mathrm{K}$,
and $\alpha = 4.44_{-0.28}^{+0.39} $.
The value $\sigma \approx 0.11 \, {\mathrm{K}}$ can be compared 
to the 1 $\sigma$ typical uncertainty in the UMass-SB
Survey of $0.4\, \mathrm{K}$ (Sanders et al. 1986).  
Figure 3 shows a comparison of this fit with the observed data.
The data from Figure 2 have been re-binned into logarithmic intervals.
The points at negative brightness temperature have been 
folded about $T = 0 \, {\mathrm{K}}$
to allow both negative and positive brightness temperatures to 
appear on a logarithmic
plot.  The fit at negative temperatures shows that the noise in the
survey deviates from a pure Gaussian: 
there is a tail to the distribution
where a few pixels trail off to brightness 
temperatures as low as $ - 1 \, \mathrm{K}$,
as well as an excess of points near $T = 0 \, \mathrm{K}$ from spectra
having better-than-specified noise.
This is to be expected in a large survey where the noise level
is not perfectly controlled.
The detected points at positive $T$ are not fit perfectly by a power
law: there are too many pixels at brightness temperatures 
near $T \approx 2.5 \,
\mathrm{K}$ and too few pixels at brightness temperatures $T > 10 \,
\mathrm{K}$ for any power law to fit.  
The best-fit power law is surprisingly steep: $N (T) \propto 
(T + 0.768 \, {\mathrm{K}})^{-4.44}$.
The large value of $C = 0.768 \, {\mathrm{K}}$ 
implies that there is no evidence for 
extensive low-level emission below the $0.1 \, {\mathrm{K}}$ noise
limit of the survey.

\section{Longitude-Velocity Maps}

We present the Bell Laboratories \tco survey in the form of
longitude-velocity maps at latitudes separated by $0^{\circ}\!.05$
(Figure 4).  The reduced FITS cubes have been smoothed with a boxcar
function to improve the signal-to-noise ratio.  The boxcar function
was $6'$ wide in longitude and $2 \, \kms$  wide in the velocity direction.
The velocity range of the observations is visible in Figure 4 as the
width in velocity of the grey dots, which are at the noise level
of the survey.
The color scale of the maps represents increasing brightness 
temperature in the
order: grey, blue, green, yellow, and red.  The color maps were
generated using the WIP graphic package.
  
Like all galactic CO surveys to date,
the data in this survey are spatially undersampled, in the
sense that the spacing between adjacent beams is more than the
half beamwidth required by the Nyquist sampling theorem.
As shown in Table 1, some of the Galactic Center Region data
are nearly fully sampled, with a $60''$ grid and a $103''$ beam;
in this region there is 
only slight aliasing of high spatial frequencies  
and no sampling error in the total flux.
The majority of the data, however,
are sampled on a grid whose spacing is two beamwidths or more;
these data do not contain information about spatial wavelengths smaller
than $6'$, and there can be substantial error in the total flux of
small sections of the survey. The maps will not accurately reproduce 
small scale structures and may contain artificial
structures because of the lack of information.
In particular, a small cloud that happened to lie between 
grid points at the $6'$ or $3'$ spacings
could be missed entirely.  Such a cloud might be as large
as $\sim 10^3$ solar masses if it were 15 kpc distant.
We have assumed in making the $l$--$v$
diagrams that the most probable value at the unobserved points is an
interpolation from the adjacent points.  Since the grid points are
arbitrarily chosen to fall at regular intervals in galactic
coordinates, they are a ``fair sample", and it is unlikely that there
is significant sampling error in the \tco flux of regions larger
than a square degree.
Care must be taken when considering small sections of the survey, 
however, because of the spatial incompleteness: data at
grid points actually observed retain all the original information as
observed, but the interpolated points could be highly inaccurate.

The velocity range of $-27$ to 147 \kms for the region $5^{\circ} < l <
45^{\circ}$ is enough to encompass the whole velocity range of
molecular gas emission, with the exception of one feature.
This elongated feature at ($l, b, v$) =
($8^{\circ}, -0^{\circ}.5, +147$ \kmsc) in Figure 4 extends 
beyond the positive
velocity limit of this survey.  The Columbia survey shows this feature
extending to 220 \kms (Dame et al. 1987). 

The following are some of the principal characteristics of the survey
($l>5^{\circ}$).

\begin{enumerate}

\item It is found that only $(1 - f_z) = $ 12\% of the pixels in 
($v,l,b$) 3-d space are responsible for the \tco emission over the
region surveyed.  This number would be even smaller for a survey which
was more extensive in $b$, or which had larger velocity coverage.
About 39\% of the total survey emissivity arises in 3-d pixels having
brightness temperature 
above $T_R^* > 1$ K ($10\, \sigma_{\mathrm{rms}}$), which
is to say:  
$${{\int_{\mathrm{1 \, K}}^{\infty}  N(T)\, T \, {\mathrm{d}} T} \over
{\int_{- \infty}^{\infty}  N(T)\, T \, {\mathrm{d}} T}} = 0.39 ~~~. $$ 
Similarly, 16\% of the survey emissivity
arises in pixels above 2 K, and 3.4\% arises in
pixels above 4K.  These values can be compared with the
UMass-SB \co survey's 17.5\% above 4 K ($10 \, \sigma_{\mathrm{rms}}$). 
 
\item A comparison with the UMass-SB survey shows that at any point in
the survey, the \tco emission is related to the \co emission.  That
relationship can be expressed as a monotonic function with scatter
(Stark et al. 1983).  Even though the \tco is roughly predictable from
the \co emission in a statistical sense, features of the clouds are
more clearly delineated in \tco than \coc. In Figure 5, an example of
\co and \tco longitude-velocity maps for the region $33^{\circ}\!.5 <
l < 36^{\circ}\!.5$ at $b = 0^{\circ}\!.1$ is shown. The \co map was
also taken using the Bell Labs 7 m telescope, but it differs only
slightly from the UMass-SB survey data.  In the \co map,
cloud boundaries are not clear, and clumps are not unambiguously
delineated. In the \tco map, clouds are more readily distinguishable.
The cloud identification using the \tco line is much easier than that
using the \co line since the \tco lines are more optically thin, so
that the cloud boundaries can be determined at a lower threshold
brightness temperature.  

\item Using the modified cloud identification code within IRAF (Lee et
al. 1997), \tco clouds have been identified and cataloged as a
function of threshold  brightness temperature.  
There are 1250 molecular clouds
identified with 1 K threshold; 560 clouds with 2 K
threshold, and 240 clouds with 3 K.  Clouds with the
hottest cores (T$_R^*$(\tco) $>$ 3 K) are restricted to the 6 kpc Molecular
Ring ($l<40^{\circ}$) and $l=80^{\circ}$ region.  It is found that the peak
brightness 
temperatures in the Galactic Center ($|l| < 5^{\circ}$) are not as
high as those in the Molecular Ring (see Figure 4, $b=0^{\circ}\!.3$ to
$-0^{\circ}\!.3$).  
In addition to cloud identification, cloud core regions can also be
located within the cloud complex. A sample spectrum centered on
($l,b$) = ($34^{\circ}\! .25, 0^{\circ}\! .1$) is presented in Figure
6.  While there are saturated and self-absorbed features in \co
emission (thin solid line), the \tco lines show clean, nearly
Gaussian, profiles.  At the location ($l,b$) = ($34^{\circ}\!.25,
0^{\circ}\!.1$) the core region is obvious in \tco map, while it
is not apparent in the \co map.  Dense core regions can be located
using this \tco survey  and then combined with existing \co data.  The
optical depth of \tco can be estimated for each channel and pixel from
the ratio of the \co and \tco brightness assuming LTE.  This estimate
gives a lower limit to the true optical depth under non-LTE
conditions.  

\item The \tco emission toward the 6 kpc Molecular Ring Region has
more contrast than does the \co emission (Sanders et al. 1986; Clemens
et al. 1986).
Note in particular that the brightest cores are concentrated in
elongated loops corresponding to the spiral arms (see Figure 4,
$20^{\circ} < l < 50^{\circ}$, $b=0^{\circ}\!.0$). 

\item Little emission occurs in the outer Galaxy; there are
few detections at negative velocities in the range $20^{\circ} < l
<80^{\circ}$, for all survey latitudes ($b = -1^{\circ}.0$ to
$1^{\circ}.0$).  The Perseus arm does appear as a distinct
feature  at $ l > 80^{\circ}$ and negative velocities.
In Figure 7, plots of the \tco integrated intensity are represented
for the whole range of velocities observed.  
Nine latitude slices have been selected for display.
The resolution in
longitude is $3'$, and the latitudes are indicated on the upper-left
corners of the plots: $1^{\circ}\!.0, 0^{\circ}\!.7, 0^{\circ}\!.5, 0^{\circ}\!.25,
0^{\circ}\!.0, -0^{\circ}\!.25, -0^{\circ}\!.5, -0^{\circ}\!.7$ and
$-1^{\circ}\!.0$.  There 
are no data in the Galactic Center region $|l| < 5^{\circ}$ for $b
= 1^{\circ}\!.0, 0^{\circ}\!.7, -0^{\circ}\!.7$ and $-1^{\circ}\!.0$. 

\item Much of the gas in the Galactic Center region ($|l| < 6^{\circ}$)
occurs at velocities forbidden to circular orbits and is likely to be
in eccentric orbits about the center (Bally et
al. 1987, Binney et al. 1991). This gas has large linewidths (see
Figure 4, $b = -0^{\circ}.6$ to $0^{\circ}.6$).  
In addition to these well-known Galactic Center features, there are
several clouds outside the Galactic Center with linewidths as large as
20 \kmsc \, that are confined to a few adjacent grid points.  Two
examples are seen in Figure 4 at ($l,b$) = ($43^{\circ}\!
.15,0^{\circ}\! .05$) and ($26^{\circ}\! .65,0^{\circ}\! .60$).   

\item Spiral structure is readily apparent in the survey from the
emission of Giant Molecular Clouds.  Within the 
volume delimited by the range of velocities permitted by galactic
rotation, there are large volumes showing these clouds (e.g. Figure 4,
$33^{\circ}<l<35^{\circ}$, $ -0^{\circ}\! .25 < b < 0^{\circ}\! .25$, $90 \,
\kms < v < 120 \, \kms$) and there 
are otherwise similar volumes within the interarm regions,  which have
no Giant Molecular Clouds (e.g. Figure 4, $35^{\circ}<l<37^{\circ}$, $
-0^{\circ}\! .25 < b < 0^{\circ}\! .25$, $85 \, \kms < v < 115 \, \kms$).

\item Small clouds are found at all permitted velocities
inwards of the solar circle.  This is not the case outside the solar
circle.  There is a complete absence of detected \tco emission in the
interarm region between the local molecular gas and the Perseus Arm
($90^{\circ} $ to $117^{\circ}$) at velocities from $-20 $ to $-40$
\kms (see Figure 4 at all values of $b$, also Cohen et al. 1980; Heyer et
al. 1998).  Cohen et al. (1980) argue that this gap 
does not necessarily mean that molecular gas is non-existent in the
interarm region; such gas could be either diffuse, cold or in
small clumps that could escape detection.  
Heyer et al. (1998) showed, however, that not even small clumps 
are detectable in \co in this region, implying a complete
absence of molecular material. 
In an interarm region somewhat closer to the Galactic Center, our
survey data show small cloudlets in the interarm region from $l =
60^{\circ}$ to $70^{\circ}$.  In the Molecular Ring at $l \sim
25^{\circ}$, the interarm regions are filled with clouds.  

\end{enumerate}

\acknowledgments

The authors wish to acknowledge the tremendous contribution of the members
of the Bell Laboratories Radio Physics Research Group during the
ten-year period of these observations: R. W. Wilson, J. Bally,
D. Mumma, W. Bent, W. D. Langer, G. R. Knapp, and M. Pound.  Special
thanks go to Jim Morgan (University of Maryland), who helped solve
a problem with WIP. This work was supported in part by Creative 
Research Fund
NZ-01-01-A-31 of Ministry of Science and Technology, Korea, and the
Internal Research and Development fund of Smithsonian Astrophysical
Observatory.

\par\noindent
\begin{deluxetable}{crrrrrrrr}
\footnotesize
\tablecaption{Coverage and Sampling of the Bell Laboratories \tco
Survey. \label{tbl-1}} 
\tablewidth{0pt}
\tablehead{
\colhead{ $l$ Range} & \colhead{$b$ Range}   & \colhead{Velocity
Range}   & \colhead{Spatial} &  \colhead{Velocity } \\ \colhead{} &
\colhead{} &   \colhead{}     & \colhead{Gridding}  &
\colhead{Resolution}
} 
\startdata
$-5^{\circ} < l < 5^{\circ}$ & $|b| < 0^{\circ}\! .6$ & $-248$ to 247 \kmsc & $6' \quad $ & 2.72 \kmsc \\
$- 0^{\circ}\! .5 < l < 0^{\circ}\! .9^{\dagger} $ & $|b|< 0^{\circ}\! .4$ & $-248$ to 247 \kmsc & $1' \quad $ & 2.72 \kmsc \\
$ 5^{\circ} < l < 45^{\circ} $ & $|b| < 0^{\circ}\! .5$ & $-27$ to 147 \kmsc & $3' \quad $ & $0.68 $ \kmsc \\
$ 5^{\circ} < l < 45^{\circ} $ & $0^{\circ}\! .5 < |b| < 1^{\circ}$ & $-27$ to 147 \kmsc & $6'
\quad $ & $0.68 $ \kmsc \\
$ 45^{\circ} < l < 113^{\circ} $ & $|b| < 0^{\circ}\! .5$ & $-87$ to 87 \kmsc & $3' \quad $ & $0.68 $ \kmsc \\
$ 45^{\circ} < l < 113^{\circ} $ & $0^{\circ}\! .5 < |b| < 1^{\circ}$ & $-87$ to 87 \kmsc & $6'
\quad $ & $0.68 $ \kmsc \\
$ 113^{\circ} < l < 117^{\circ} $ & $|b| < 0^{\circ}\! .5$ & $-127$ to 47 \kmsc & $3' \quad $ & $0.68 $ \kmsc \\
$ 113^{\circ} < l < 117^{\circ} $ & $0^{\circ}\! .5 < |b| < 1^{\circ}$ & $-127$ to 47 \kmsc & $6' \quad $ & $0.68 $ \kmsc \\
\enddata
\tablenotetext{}{$^{\dagger}$ This region was reobserved with
higher sampling rate of $1'$ (see Bally et al. 1987).}

\end{deluxetable}

\clearpage

\centerline{FIGURE CAPTIONS}

\par\noindent
Figure 1.  Beamsizes and spatial sampling grids of major CO galactic plane
surveys.  \co surveys of UMass-SB
(beamsize $45''$) and Columbia survey ($8.'7$) are represented in
solid-line circles, and \tco surveys of Nagoya University ($170''$) and
Bell Laboratories ($103''$) are represented as thick solid-line circles
on a $3'$-grid.

\par\noindent
Figure 2. 
Distribution of \tco brightness temperatures in all channels of 
the spectra of the whole region surveyed is represented.  The logarithmic
scale of the histogram is represented in the inset. 

\par\noindent
Figure 3. 
Parametric fit to the distribution of brightness 
temperatures in the Bell Labs \tco
Survey.  The abscissa is the log of the absolute value of brightness
temperature in survey pixels, in Kelvin.  The ordinate is the log of
the number of survey pixels having that brightness temperature.
The points show the number of pixels in a small range of brightness
temperatures
divided by the width of that temperature range.
The open circles are at negative brightness temperature.  Almost all
of the negative pixels actually have zero emission and appear at
negative brightness temperature because of noise in the survey.
Filled circles are at
positive brightness temperature.  
The solid line shows a parametric model of
the distribution function.

\par\noindent
Figure 4. 
 Longitude-velocity color scale maps at all latitude grid points.
The color map represents increasing intensity
as grey, blue, green, yellow, and red.
The velocity range of the observations is delimited by 
grey dots representing the noise in the data.

\par\noindent
Figure 5. An example of longitude-velocity \tco and \co contour maps
at $ l = 33^{\circ}\!.5$ to $36^{\circ}\!.5$, and $b = 0^{\circ}\!.1$. The
\co contours are multiples of 1.0 K up to a 
maximum contour of 12 K.  The \tco contours are 0.3, 0.6, 1.0 K, followed by
increments of 0.6 K, up to a maximum contour of 6.4 K.

\par\noindent
Figure 6. \co and \tco composite spectra centered on and around
($l,b$) = ($34^{\circ}\!.25, 0^{\circ}\!.1$).  The thin solid line
represents \coc, and   the thick solid line represents \tco.

\par\noindent 
Figure 7.  Plots of the \tco intensity integrated over
the whole range of velocity, plotted against longitude at the
indicated latitude, $1^{\circ}\! .0, 0^{\circ}\! .7, 0^{\circ}\! .5,
0^{\circ}\! .25, 0^{\circ}\! .0, -0^{\circ}\! .25, -0^{\circ}\! .5,
-0^{\circ}\! .7$ and $-1^{\circ}.0$. The resolution in longitude is
$3'$.

\end{document}